\documentclass[aps,preprintnumbers,amsmath,amssymb,nofootinbib]{revtex4}  
\usepackage{graphicx}
\usepackage{bm}
\usepackage{epsfig}
\usepackage{color}
\usepackage{float}

\begin{document}
\title{
Non-factorizable charming loops in FCNC B decays
\\
vs B-decay semileptonic form factors}
\author{
Dmitri Melikhov$^{a,b,c}$}
\affiliation{
$^a$D.~V.~Skobeltsyn Institute of Nuclear Physics, M.~V.~Lomonosov Moscow State University, 119991 Moscow, Russia\\
$^b$Joint Institute for Nuclear Research, 141980 Dubna, Russia\\
$^c$Faculty of Physics, University of Vienna, Boltzmanngasse 5, A-1090 Vienna, Austria}
\date{\today}
\begin{abstract}
We compare a non-factorizable charming-loop correction to an exclusive FCNC $B$-decay 
given in terms of the 3-particle Bethe-Salpeter amplitude (3BS) of the $B$-meson,
$\langle 0|\bar q(x)G_{\mu\nu}(z)b(0)|B(p)\rangle$,   
with the corresponding correction to the $B$-meson semileptonic form factor. 
In spite of certain similarities, these two corrections are shown to have substantial differences:
The form factor correction is dominated by the {\it collinear} light-cone configuration of 3BS: 
$z_\mu= u x_\mu$, $0 < u < 1$, $x^2=0$. In contrast, the FCNC amplitude is dominated by a different
configuration with {\it non-collinear} arguments: $x^2=0$, $z^2=0$, but $(x-z)^2\ne 0$ (i.e., $z_\mu\ne u x_\mu$). 
\end{abstract}
\maketitle

\section{Introduction}
\label{Sec_introduction}
Charming loops in rare FCNC (flavour-changing neutral current) decays of the $B$-meson 
have impact on the $B$-decay observables \cite{neubert} providing an unpleasant noise for
the studies of possible new physics effects (see, e.g., recent discussions
\cite{ciuchini2020,ciuchini2021,diego2021,matias2022} and refs therein).  

A number of theoretical analyses of non-factorizable (NF) charming loops in FCNC $B$-decays has
been published. We mention here those directly related to the discussion of this paper: 
in \cite{voloshin}, an effective gluon-photon local operator describing the charm-quark loop has been 
calculated as an expansion in inverse charm-quark mass $m_c$ and applied to inclusive $B\to X_s\gamma$ decays
(see also \cite{ligeti,buchalla}); in \cite{khod1997}, NF corrections in $B\to K^*\gamma$ using local
operator product expansion (OPE) have been studied;
NF corrections induced by the {\it local} photon-gluon operator 
have been calculated in \cite{zwicky1,zwicky2} in terms of the light-cone (LC) 3-particle antiquark-quark-gluon
Bethe-Salpeter amplitude (3BS) of $K^*$-meson \cite{braun,ball1,ball2} with two field operators having equal coordinates,  
$\langle 0| \bar s(0)G_{\mu\nu}(0) u(x)|K^*(p)\rangle$, $x^2=0$. However, local OPE for the charm-quark loop in FCNC $B$ decays
leads to a power series in $\Lambda_{\rm QCD} m_b/m_c^2$; numerically this parameter is close to one.
To sum up $O(\Lambda_{\rm QCD} m_b/m_c^2)^n$ corrections, Ref.~\cite{hidr} obtained a {\it nonlocal} photon-gluon
operator describing the charm-quark loop and evaluated its effect making use of 3BS of the $B$-meson
in a collinear LC approximation $\langle 0| \bar s(x)G_{\mu\nu}(ux) b(0)|B(p)\rangle$, $x^2=0$. 
This approximation was used later for the analysis of other FCNC $B$-decays \cite{gubernari2020}.

The collinear LC configuration is known to provide the dominant 3BS contribution to meson 
form factors \cite{braun1994,offen2007}, in particular, to form factors of semileptonic (SL) $B$-decay
induced by the tree-level $b\to u$ weak charged current (CC). So it may seem attractive to express also the
FCNC $B$-decay amplitude via this collinear LC 3BS of the $B$-meson.

However, the 3BS contribution to the CC $B$-decay and to the FCNC $B$-decay have a qualitative difference: 
Let us consider the $B$-decay in the $B$-meson rest frame.
In CC $B$-decays, the $b$-quark emits a fast light $u$-quark which is later hit by a soft gluon and thus
keeps moving in the {\it same} space direction. In FCNC $B$-decays, a fast light $s$-quark and a pair of
fast $c$-quarks emitted by the $b$-quark move in the {\it opposite} space
directions. We shall demonstrate that, as the consequences of this difference,
the $B$-meson CC weak form factor is dominated by a {\it collinear} LC configuration 
$\langle 0| \bar q(x)G_{\mu\nu}(ux) b(0)|B(p)\rangle$, $x^2=0$ \cite{japan}, whereas the FCNC $B$-decay amplitude 
is dominated by a {\it non-collinear} configuration $\langle 0| \bar q(x)G_{\mu\nu}(z) b(0)|B(p)\rangle$,
$x^2=0$, $z^2=0$, but $(x-z)^2\ne 0$ \cite{mk2018,m2019}. The first application of a NC 3BS 
to FCNC $B$-decays was presented in \cite{wang2022}. 

We study the general properties of the 3BS contributions to the amplitudes of $B$-decays and
formulate the conditions necessary for the dominance of the amplitude by a collinear 3BS configuration. We
perform the analysis using field theory with scalar quarks/gluons which is free of technical complications
and allows one to focus on the conceptual issues; the generalization of our analysis to QCD is straightforward.
Section~\ref{sect:cc_vs_fcnc} demonstrates the technical similarities between the CC and the FCNC
amplitudes and their equivalence to the generic diagram of the form factor topology, and Section~\ref{sect:collinear}
studies the conditions under which this generic diagram is dominated by a collinear 3BS configuration.
As follows from this analysis, large $O(1)$ corrections to the collinear LC 3BS contribution should emerge
in the amplitudes of FCNC $B$-decays.  
Chapter~\ref{sect:fcnc} studies in detail the FCNC $B$-decay amplitude, including the cases of the
light $u$-quark and the $c$-quark in the triangle loop, adopting for the latter case the counting
scheme $\Lambda_{\rm QCD}m_b/m_c^2\simeq 1$ \cite{paz}.
The origin of large $O(1)$ corrections to the collinear LC approximation is identified: namely, 
we show that a non-collinear 3BS configuration (both $x$ and $z$ on the light cone, but on different axes:  
$x$ on the $(+)$ axis, and $z$ on the $(-)$ axis or vice versa  \cite{paz2010}) give parametrically unsuppressed contributions
compared to the collinear LC 3BS contribution.
So, the full dependence of 3BS of the $B$-meson on the variable $(x-z)^2$ is necessary
to sum properly the $(\Lambda_{\rm QCD}m_b/m_c^2)^n$ corrections in FCNC $B$-decays. 

\section{The amplitudes of $B$-meson CC decay vs FCNC decay}
\label{sect:cc_vs_fcnc}
In this Section we show that the 3BS contributions to the CC and to the FCNC amplitudes may be reduced to the diagram 
of a generic form factor topology with an (essential) difference of the location of the heavy-quark field in this diagram. 
\subsection{The amplitude of semileptonic $B$-meson decay induced by weak charged current}
\begin{figure}[h]
  \includegraphics[height=4cm]{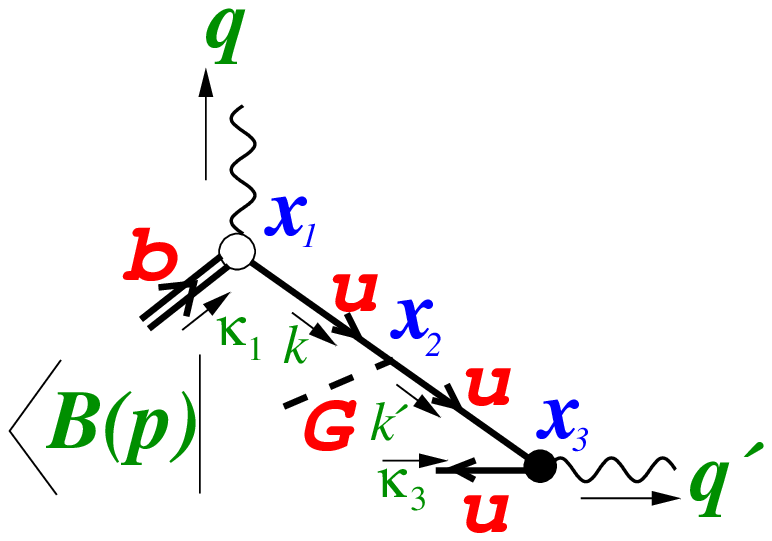}
  \caption{\label{Fig1} Momentum notations in the 3BS contribution to the form factor describing weak $b\to u$ semileptonic
    $B$-decay. }
\end{figure}

To exemplify the essential part of our analysis, we neglect the spins of the $B$-meson constituents
(quarks and gluons are treated as scalar fields) as well as the Lorentz structure of the weak currents.
So, instead of the full QCD amplitude describing a semileptonic (SL) $B$-meson weak decay induced by a charged current,
\begin{eqnarray}
  \label{Acc0}
  A^{\nu}_{\rm SL}(p|q,q')=
  i\int dx_1 dx_3e^{i q x_1+iq'x_3}
  \langle 0|T\left\{\bar u(x_3){\cal O} u(x_3),\bar u(x_1)\gamma^\nu(1-\gamma_5)b(x_1)\right\}|B_u(p)\rangle, 
\end{eqnarray}
$\cal O$ is a Dirac matrix, we consider the amplitude 
\begin{eqnarray}
  \label{Acc1}
  A_{\rm SL}(p|q,q')=
  i\int dx_1 dx_3 e^{i q x_1+iq'x_3}
  \langle 0|T\left\{u^\dagger(x_3)u(x_3),u^\dagger(x_1)b(x_1)\right\}|B_u(p)\rangle 
\end{eqnarray}
with scalar ``quarks'' and ``gluons''. Here $q$ is the momentum emitted by the weak vertex,
and $q'$ is the momentum emitted by the interpolating current of the outgoing state, see Fig.~\ref{Fig1}. 
The $u$ and the $b$ quark fields are the Heisenberg operators with respect
to the strong interaction. We will be interested in the part of the amplitude (\ref{Acc1}) that emerges at the
first order of the expansion in the strong coupling and contains the gluon field $G$. The corresponding Feynman
graph is shown in Fig.~\ref{Fig1}. After some manipulations
(and making use of the Fock-Schwinger gauge, see \cite{braun1994,offen2007})
the relevant to us part of this amplitude may be written as
\begin{eqnarray}
 \label{Acc2}  
A_{\rm SL}(p|q,q')&=&\int dx_1 dx_2 dx_3 dk  dk'
\,e^{i q x_1 -i k(x_2-x_1)} \,e^{-i k'(x_3-x_2)+iq'x_3}\nonumber\\
&&\;\;\times \frac{1}{m_u^2-k^2} \frac{1}{m_u^2-k'^2}\langle 0|u^\dagger(x_3)G(x_2)b(x_1)|B_s(p)\rangle.
\end{eqnarray}
Making use of the transformation properties of field operators under translations, we may shift the coordinate
of the gluon field to zero and introduce the new variables through the relations
$k=\kappa_1-q$ and $k'=q'-\kappa_3$, $x_1-x_2\to x_1$ and $x_3-x_2\to x_3$. 
After that, we perform the $x_2$-integration that leads to the momentum conservation $\delta(p-q-q')$,
and the amplitude (\ref{Acc2}) takes the form  
\begin{eqnarray}
A_{\rm SL}(p|q,q')&=&(2\pi)^4\delta(p-q-q')\overline A_{\rm SL}(p|q,q'), 
\end{eqnarray}
with 
\begin{eqnarray}
 \label{Acc3}  
\overline A_{\rm SL}(p|q,q')&=&
 \int dx_1 dx_3 d\kappa_1 d\kappa_3
 \,e^{i \kappa_1 x_1 +i \kappa_3 x_3}
 \frac{1}{m_u^2-(\kappa_1-q)^2} \frac{1}{m_u^2-(q'-\kappa_3)^2}\langle 0|u^\dagger(x_3)G(0)b(x_1)|B_s(p)\rangle.
\end{eqnarray}

\subsection{Non-factorizable part of the amplitude of FCNC $B$-decay }
\begin{center}
\begin{figure}[h]
\includegraphics[height=5cm]{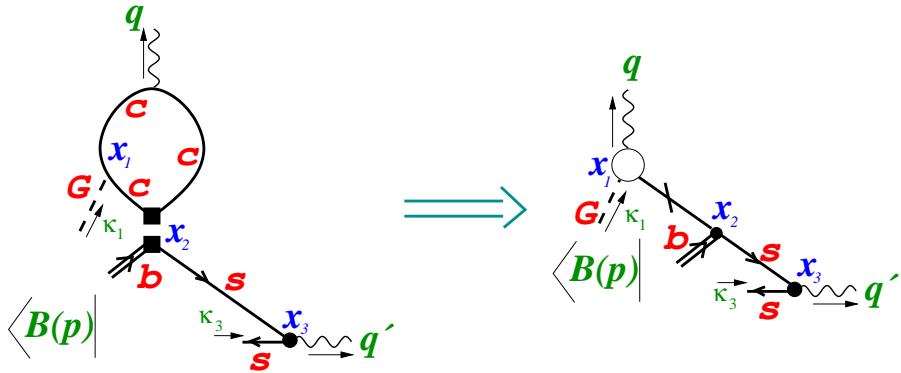}
\caption{\label{Fig2}Feynman diagram describing the 3BS contribution to a NF amplitude of FCNC $B$-decay.
The crossed propagator line means that the propagator is replaced by the triangle
charming loop $\Gamma_{cc}(\omega_1p,q)$.}
\end{figure}
\end{center}

The amplitude of FCNC $B_s$-decay such as, e.g. $B_s\to \gamma^*\gamma^*$ decay, is given by the
following expression \cite{mnk2018}: 
\begin{equation}
\label{AFCNC1}
A^{\nu\mu}_{\rm FCNC}(p|q,q')= 
\int dx \exp(i q x) dx_3 \exp(i q'x_3)
\langle 0|T\left\{\bar c(x)\gamma^\nu c(x), \bar s(x_3)\gamma^\mu s(x_3)\right\}|B_s)p)\rangle. 
\end{equation}
Here the $c$ and $s$ quark operators are the Heisenberg operators with respect to strong and weak interactions.
Notice that in the case of an FCNC $B$-decay, the $b$-quark field is not contained in the current
operators under the $T$-product, but comes into the came at order $G_F$ in weak interaction. The part of the amplitude describing the non-factorizable
contribution of the charming loop emerges at order
$G_F$ in weak interaction and at the first order in strong interaction. The contribution of the charming loop
is given through the gluon field $G_{\mu\nu}$ \cite{lm}, so the Feynman diagram describing the FCNC amplitude (\ref{AFCNC1})
is represented by Fig.~\ref{Fig2}.
We omit all complications related to Lorentz and spinor structure details and consider scalar quark
fields $b$, $c$, and $s$, and a scalar gluon field $G$. We then come to the following expression for
the non-factorizable part of the amplitude of FCNC $B$-decay, see Fig.~\ref{Fig2}: 
\begin{eqnarray}
  A_{\rm FCNC}(p|q,q')=(2\pi^4)\delta(p-q-q')\overline A_{\rm FCNC}(p|q,q')
\end{eqnarray}
with
\begin{eqnarray}
\overline A_{\rm FCNC}(p|q,q')= \frac{G_F}{\sqrt{2}}\int dx_1 dx_3 d\kappa_1 d\kappa_3
  e^{i x_1 \kappa_1 +i\kappa_3 x_3}
  \Gamma_{cc}(\kappa_1, q)
  \frac{d\kappa_3}{m_s^2-(q'-\kappa_3)^2}
\,\langle 0| s^\dagger(x_3)G(x_1) b(0)|B_s(p)\rangle.
\end{eqnarray}
Here $\Gamma_{cc}(\kappa_1,q)$ is the charm-quark triangle diagram which
may be written as a double integral in Feynman parameters
(see a detailed discussion in \cite{hidr,mk2018,m2019,mnk2018,lm})
\begin{eqnarray}
\label{gammacc1}
\Gamma_{cc}(\kappa_1, q)=\frac{1}{8\pi^2}
\int\limits_{0}^{1}du \int\limits_{0}^{1}dv 
\frac{\theta(u+v<1)}{m_c^2-uv (\kappa_1-q)^2 -u(1-u-v)\kappa_1^2-v(1-v-u)q^2}. 
\end{eqnarray}
Important for us is that the quantity is a quadratic function in momentum variables. So, the FCNC amplitude is similar to
the SL amplitude, with the light-quark propagator replaced by an ``effective'' propagator $\Gamma_{cc}(\kappa_1, q)$
\begin{eqnarray}
\frac{1}{m_u^2-(\kappa_1-q)^2}\to \Gamma_{cc}(\kappa_1, q). 
\end{eqnarray}
The main difference between the SL and the FCNC amplitudes arises from the fact
that the heavy $b$-quark in a SL decay amplitude is attached to 
the {\it end-point} of the line connecting $x_1$ and $x_3$, along which the energetic light quarks are propagating,
while in the FCNC the heavy field is attached to the {\it middle point} of the line connecting $x_1$ and $x_3$.
We shall see that this features of the SL and FCNC $B$-decays are responsible for a qualitative difference
between the configurations of the 3BS of the
$B$-meson, that provide the dominant contributions in SL $B$ decays and in FCNC $B$-decays. 

To demonstrate this difference, the next Section considers the general amplitude of the
form-factor topology and figures out the properties necessary for the dominance of the collinear 3BS configuration.

\newpage
\section{3BS contribution to a generic amplitude of the form factor topology
\label{sect:collinear}}
\begin{center}
\begin{figure}[h!]
\includegraphics[height=4.7cm]{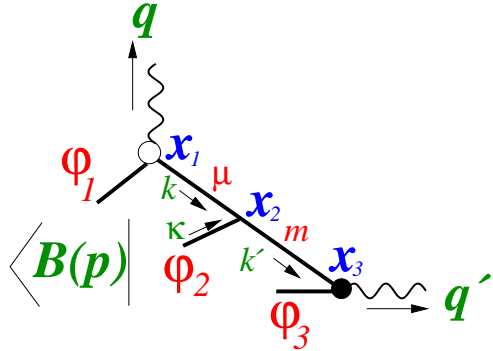}
\caption{\label{Fig3} 3BS contribution to the generic amplitude of the form factor topology.}
\end{figure}
\end{center}

Let us consider the generic form factor amplitude $A(p|q,q')$ shown in Fig.~\ref{Fig3}.
As we have mentioned above, both $A_{\rm FCNC}$ and $A_{\rm SL}$ are reduced to this amplitude, such that
each of the FCNC and the SL amplitudes is characterized by a specific (and different) content of the heavy and the light
fields $\varphi_{1,2,3}$: in FCNC decays, the field $\varphi_{2}$ is heavy, while $\varphi_{1,3}$ are light;
in SL decays, $\varphi_{1}$ is heavy, while $\varphi_{2,3}$ are light. If one considers the case of the 3BS correction to the
form factor of a light meson, all fields $\varphi_{1,2,3}$ are light degrees of freedom (light quarks or gluons). 

The properties of the set of meson constituents $\varphi_{1,2,3}$ (i.e.,
which of these fields are heavy and which are light) are reflected
in the properties of the amplitude $\langle 0|\varphi_1(x_1)\varphi_2(x_2)\varphi_3(x_3)|B(p)\rangle$.
The goal of our analysis in this Section is to figure out the kinematical configuration of the constituent
fields that dominate the amplitude of Fig.~\ref{Fig3}. 

The analytic expression corresponding to the diagram of Fig.~\ref{Fig3} has the form 
\begin{eqnarray}
\label{A1}
A(p\,|q,q')=\int 
\frac{dx_1\,dx_2\, dx_3\, dk\, dk'}{(\mu^2-k^2)(m^2-k'^2)}e^{i q x_1 -i k(x_2-x_1)-ik'(x_3-x_2)+i q'x_3}
\langle 0|\varphi_1(x_1)\varphi_2(x_2)\varphi_3(x_3)|B(p)\rangle.  
\end{eqnarray}
As already mentioned above, the amplitude contains $\delta(p-q-q')$ which may be isolated by
making use of the transformation properties of the
field operators $\varphi_{1,2,3}$ under translations: one can set one of the arguments of the field operators equal zero,
and integrate over one of the coordinate differences.
For the moment we will not make use of this property and will keep all three arguments
$x_{1,2,3}$ nonzero.

By introducing the Feynman parameter $v$ to combine two propagators in a single propagator squared, and 
after redefinitions of the variables
\begin{eqnarray}
&&  \tilde k=k-v\kappa, \\
&&  x_2=x_1(1-v)+x_3 v +z_2, 
\end{eqnarray}
the amplitude takes the form convenient for a further analysis
\begin{eqnarray}
\label{A2}
A(p|q,q')&=&\int\limits_0^1 dv \int d \kappa \,d\tilde k\, dx_1\, dz_2\,dx_3\
\frac{e^{i x_1(\tilde k + q)+ ix_3(q'-\tilde k)+i \kappa z_2}}
{\left[m^2(1-v)+\mu^2 v -\tilde k^2-v(1-v)\kappa^2\right]^2}\nonumber\\
&&\times\langle 0|\varphi_1(x_1)\varphi_2(x_1(1-v)+x_3 v+z_2)\varphi_3(x_3)|B(p)\rangle, 
\end{eqnarray}
Here $\kappa$ is the momentum transferred in the {\it central} point $x_2$ and the variable $z_2$
measures the deviation of the configuration $x_1,x_2,x_3$ from the straight line joining the
end points $x_1$ and $x_3$.

To proceed further, one can attempt to expand
$\langle 0|\varphi_1(x_1)\varphi_2(x_1(1-v)+ x_3 v+z_2)\varphi_3(x_3)|B(p)\rangle$
in powers of $z_2$ and obtain  in this way a tower of collinear operators of the increasing
dimension containing derivatives of $\varphi_2$. The expansion in powers of $z_2$ corresponds to expanding the denominator
in powers of $\kappa^2$. Such expansion is meaningful if $\kappa$ is soft compared to the virtualities of the
propagators $D(x_1-x_2)$ and $D(x_2-x_3)$. 
Obviously, the amplitude is dominated by a collinear configuration of the 3BS
{\it only} if the momentum transferred in the central vertex is soft 
compared to the virtualities of the particle propagators along the line $x_1$-$x_3$.

There are phenomenologically relevant cases where the collinear 3BS configuration
indeed dominates the amplitude $A(p|q,q')$:
\begin{itemize}
\item[$\bullet$]
QCD radiative correction to the $B\to j_1 j_2$ weak form factor.
In this case the field $\varphi_1$ is heavy, whereas $\varphi_2$ and $\varphi_3$ are light
(the gluon and the light quark, respectively). If
$q^2, q'^2 \ll M_B^2$, $\kappa^2=O(\Lambda_{\rm QCD}^2)$, the virtualities of the particles propagating 
along the segments $x_1-x_2$ and $x_2-x_3$ are $O(m_b\Lambda_{\rm QCD})$, and $\kappa^2\simeq \Lambda_{\rm QCD}^2$.
The expansion in $\kappa^2$ seems meaningful and the amplitude is dominated by a collinear light-cone (LC) configuration
$x_2=x_1\,(1-v)+x_3\,v$, $x_1^2\simeq 0$ and $x_3^2\simeq 0$. 
\item[$\bullet$]
QCD radiative correction to $M\to j_1 j_2$ form factor, $M$ is a light meson.
In this case, all three fields $\varphi_{1,2,3}$ are light ($\varphi_1$ and $\varphi_3$ are the light
quarks and $\varphi_2$ is the gluon). To make the OPE convergent, we cannot set $q^2=q'^2=0$, but must keep
$q^2\simeq q'^2 \ll -1\,{GeV}^2$. In this case both quark propagators are
highly virtual, $O(q^2,q'^2)$, the momentum transfer $\kappa$ is soft, $\kappa^2=O(\Lambda_{\rm QCD}^2)$,
and the collinear 3BS configuration dominates the amplitude.
\end{itemize}
Unfortunately, the non-factorizable correction to the FCNC amplitude of the $B$-meson decay does
not fall into this class of processes:
$\varphi_1$ and $\varphi_3$ are light degrees of freedom (the gluon and the light quark, respectively), whereas
$\varphi_2$ is a heavy quark which carries almost the full momentum of the $B$-meson, $\kappa_2^2\sim M_B^2$. 
The momentum $\kappa_2$ is thus by far not soft compared to the virtualities of the particles along the
line $x_1$-$x_3$, and the expansion around the collinear 3BS configuration does not converge:
Expanding in powers of $\kappa_2^2$ leads to a series in which all terms have the same order of magnitude. 

In the next Section, we look in more detail what kind of expansion of the amplitude arises in this case.
In particular, we show that a non-collinear 3BS configuration with 
$(x_1-x_2)^2=0$, $(x_2-x_3)^2=0$, but $(x_1-x_3)^2\ne 0$ dominates the FCNC amplitude.



\section{The amplitude of FCNC $B$-decay
\label{sect:fcnc}}
We start with briefly recalling the general properties of the 3-particle (antiquark-quark-gluon) BS
wave function of the meson and then obtain the $B$-decay FCNC amplitude in terms of this BS wave function.
To simplify the analytical expressions, we will consider the case $q^2=q'^2=0$. 

\subsection{Parametrization of the 3-particle BS amplitude}
The 3-particle BS amplitude may be written in the form (see e.g. \cite{braun1994,ball2,japan})\footnote{
Refs.~\cite{braun1994,ball2,japan} provide the expansion of 3BS for collinear arguments $x_1=v x_3$.
A generalization to the case of non-collinear arguments presented here is straightforward.}
\begin{eqnarray}
  \label{3da}
\langle 0|G(x_1)b(x_2)s^\dagger(x_3)|B_s(p)\rangle=
\int D \omega e^{-i \omega_1 x_1 p -i \omega_2 x_2 p-i \omega_3 x_3 p}
\left[\psi_0(\omega)+\psi_{12}(\omega)x_{12}^2+\psi_{13}(\omega)x_{13}^2+\psi_{23}(\omega)x_{23}^2+\dots \right],
\nonumber\\
\end{eqnarray}
where 
\begin{eqnarray}
x_{ij}^2=(x_i-x_j)^2,\quad D\omega\equiv d\omega_1 d\omega_2 d\omega_3 \delta(1-\omega_1-\omega_2-\omega_3), \quad 
\psi_i(\omega)\equiv \psi_i(\omega_1,\omega_2,\omega_3). 
\end{eqnarray}
Since the $b$-quark is heavy, all functions $\psi_i(\omega)$ in the amplitide (\ref{3da}) 
have support in the end-point regions 
\begin{eqnarray}
\omega_2\sim 1-O(\Lambda_{\rm QCD}/m_b),\quad &&\omega_{1,3}\sim O(\Lambda_{\rm QCD}/m_b).
\end{eqnarray}
One of the arguments of the field operators may be set to zero by using the transformation properties of
the field operators under translations, so we set the coordinate of the heavy $b$-quark to zero $x_2=0$ (in this way 
$\delta(p-q-q')$-function describing the momentum conservation has been singled out from $A_{\rm FCNC}$).  
As the next step, we insert this expression in the general formula for the amplitude
(\ref{A1}) which takes the following form 
\begin{eqnarray}
\label{A3}
\overline A(p\,|q,q')=\int 
dx_1\, dx_3\, d\kappa_1\, d\kappa_3
e^{i \kappa_1 x_1+i\kappa_3 x_3}\Gamma_{cc}(\kappa_1,q)D_s(\kappa_3-q')\Psi_p(x_1,x_3), 
\end{eqnarray}
where
\begin{eqnarray}
\Psi_p(x_1,x_3)\equiv  \langle 0|G(x_1)b(0)s^\dagger(x_3)|B(p)\rangle
\end{eqnarray}
and $D_s(k)=1/(m_s^2-k^2-i0)$. We will not introduce the Feynman parameter $v$ to combine the propagators
as was done in Eq.~(\ref{A2}), but evaluate the amplitude directly. Following the results of the previous section,
we are going to demonstrate directly that the FCNC amplitude is indeed not dominated by
the collinear field configuration.

\subsection{The contribution of the $\psi_0$ term in 3BS (neglecting all powers of $x_{ij}^2$).}
\begin{center}
\begin{figure}[h!]
\includegraphics[height=4.5cm]{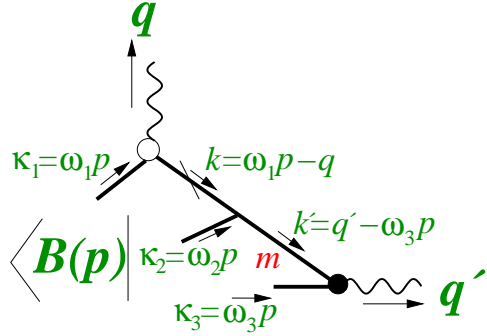}
\caption{\label{Fig3b}
  Momenta values in the 3BS contribution to the amplitude of FCNC $B$-decay.
  The crossed propagator line means that the propagator is replaced by the triangle charming loop $\Gamma_{cc}(\omega_1p,q)$}
\end{figure}
\end{center}

The term $\propto\psi_0(\omega)$ in 3BS does not contain $x_{ij}^2$ so its contribution to 
$A(q,p)$ is calculated easily: the integrals over $x_{1,3}$ give the $\delta(\kappa_1-\omega_1 p)$ and
$\delta(\kappa_3-\omega_3 p)$ so one ends up with the following expression (see Fig.~\ref{Fig3b}): 
\begin{eqnarray}
\label{Aqp}
A_{\psi_0}(q,p)=\int\limits_0^1 d\omega_1 \int\limits_0^{1-\omega_1} d\omega_3 \,\psi_0(\omega_1,\omega_3)
\Gamma_{cc}\left(\omega_1 p, q \right)D_s\left((q'-\omega_3 p)^2\right). 
\end{eqnarray}

\subsubsection{$s$-quark propagator}
The denominator of the $s$-quark propagator takes the form  
\begin{eqnarray}
m_s^2-(\omega_3 p-q')^2=m_s^2-\lambda q^2-(1-\omega_3)q'^2+(1-\omega_3)\omega_3 M_B^2. \nonumber 
\end{eqnarray}
For $q^2=q'^2=0$, in the region $\omega_3\sim \Lambda_{\rm QCD}/m_b$ that dominates the integral, the $s$-quark is highly virtual:
its momentum squared is $O(\Lambda_{\rm QCD}m_b)$. 

\subsubsection{Charm-quark loop}
The analytic expression for the triangle charming loop was already given in (\ref{gammacc1}).
We now rewrite it in a slightly different way by introducing a new variable $0<\tau<1$, $v=(1-u)\tau$: 
\begin{eqnarray}
\label{gammacc2}
\Gamma_{cc}(\kappa_1, q)=\frac{1}{8\pi^2}
\int\limits_{0}^{1}du (1-u)\int\limits_{0}^{1}d\tau  
\frac{1}{m_c^2-u(1-u)[\tau(\kappa_1-q)^2+\kappa_1^2(1-\tau)]-q^2(1-u)^2\tau(1-\tau)}. 
\end{eqnarray}
To shorten the formulas, we set hereafter $q^2=0$. This expression appears under the convolution
Eq.~(\ref{A3}) with the 3BS of the $B$-meson. 
Then the $\omega_1$-integral is peaked near $\omega_1\sim \Lambda_{\rm QCD}/m_b$ so the gluon is soft: 
$\kappa_1=\omega_1 p$ and $\kappa_1^2\sim O(\Lambda_{\rm QCD}^2)\ll m_c^2$.
In (\ref{gammacc2}), $\kappa_1^2$ may be neglected compared to $m_c^2$, and we find
\begin{eqnarray}
\Gamma_{cc}(\kappa_1,q)=\frac{1}{8\pi^2}\int\limits_0^1 du (1-u) \int\limits_0^1 d\tau
\frac{1}{m_c^2-u(1-u)\tau(\kappa_1-q)^2}.
\end{eqnarray}
Taking the $\tau$-integral one comes to a relation very similar to the one obtained
in \cite{hidr} (up to the appropriate changes related to the spins of the constituents and the Lorentz
structures of the currents).  

Let us make two remarks:
\begin{itemize}
\item
The obtained expression Eq.~(\ref{Aqp}) for the amplitude $A_{\psi_0}(q,p)$ 
does not ``feel'' the relative location of the arguments $x_i$.
In particular, $\psi_0$ is precisely the same function that parameterizes, e.g., the collinear LC
configuration discussed in \cite{hidr}. Moreover, up to technical complications related to the spins of the
$B$-meson constituents and the Lorentz structure of the currents,
the obtained expression corresponds to the approximation considered in \cite{hidr}.
So one may say that the $A_{\psi_0}$ approximation to the FCNC amplitude corresponds to the contribution of the
collinear LC 3BS of the $B$-meson. 
\item
We have shown in the previous Section, that the collinear approximation does not dominate the FCNC amplitude,
and expanding the amplitude near the collinear configuration should lead to sizeable $O(1)$ corrections to the
collinear approximation. The 3BS feels the relative location of the arguments $x_i$ only starting with the
terms $x_{ij}^2$. So, based on the general argument, one expects sizeable corrections, of the order of unity,
coming from powers of $x_{ij}^2$.
We shall see that indeed large $O(1)$ corrections emerge from all powers of $(x-z)^2$, whereas terms containing
powers of $x^2$ and $z^2$ lead to the suppressed contributions. 
\end{itemize}

\subsection{Contributions induced by $x_{ij}^2$ terms in the 3-particle BS amplitude}
We now turn to the calculation of the contributions of $x_{ij}^2$ terms in the 3BS amplitude.

In the problem under consideration one encounters two heavy quark scales, $m_c$ and $m_b$,
such that $\Lambda_{\rm QCD}\ll m_c\ll m_b$. Taking into account the real values of the quark
masses one encounters a new parameter of order of unity:
\begin{eqnarray}
\Lambda_{\rm QCD }m_b/m_c^2\simeq 1.
\end{eqnarray}
One needs therefore to sum all powers of the parameter $\Lambda_{\rm QCD}m_b/m_c^2$.
We shall see that this task is related to a summation of all corrections of the order $(x_1-x_3)^{2n}$. 

\noindent 
$\bullet$ Let us start with the $x_{1\mu}$ term under the integral.
It may be written as
\begin{eqnarray}
  x_{1\mu}e^{i \kappa_1 x_1}\Gamma_{cc}(\kappa_1,q)d\kappa_1=
-i\frac{\partial}{\partial \kappa_1^\mu}e^{i \kappa_1 x_1}\Gamma_{cc}(\kappa_1,q)d\kappa_1= 
  i e^{i \kappa_1 x_1}\frac{\partial}{\partial \kappa_1^\mu}\Gamma_{cc}(\kappa_1,q)
  \to i \frac{\partial}{\partial \kappa_1^\mu}\Gamma_{cc}(\kappa_1,q)|_{\kappa_1=\omega_1p}.
\end{eqnarray}
where we have performed the parts integration. The $x_1$-dependence then remains only in the exponential factors, 
and $x_1$-integartion may be taken and leads to $\delta(\kappa_1-\omega_1 p)$. 

\noindent $\bullet$
Similarly, the $x_{3\mu}$ term under the integral may be handled leading to $\delta(\kappa_3-\omega_3 p)$: 
\begin{eqnarray}
  x_{3\mu}e^{i \kappa_3 x_3}D_s(\kappa_3-q')d\kappa_3=
-i\frac{\partial}{\partial \kappa_3^\mu}e^{i \kappa_3 x_3}D_s(\kappa_3-q')d\kappa_3= 
  i e^{i \kappa_3 x_3}\frac{\partial}{\partial \kappa_3^\mu}D_s(\kappa_3-q')
  \to i \frac{\partial}{\partial \kappa_3^\mu}D_s(\kappa_3-q')|_{\kappa_3=\omega_3 p}.
\end{eqnarray}
Making use of these formulas we obtain the factors in the integrands
that describe the relative contributions to $A_{\psi_{ij}}(p,q)$ of the $x_{ij}^2\psi_{ij}$ terms
compared to the $\psi_{0}$ contribution to $A_{\psi_0}(p,q)$
(i.e., the factors given below are to be compared with $\psi_o$): 
\begin{eqnarray}
  \label{x11c}
x_1^2\Lambda_{\rm QCD}^2\psi_{12}:&& 
  \Lambda_{\rm QCD}^2\psi_{12}\frac{8u^2(1-u)^2(\kappa_1-q \tau)^2}
         {(\left[m_c^2-u(1-u)[(\kappa_1-q)^2\tau+\kappa_1^2(1-\tau)]\right]^2}
         \big|_{\kappa_1=\omega_1 p}\sim 
         \frac{\Lambda_{\rm QCD}^2\omega_1 M_B^2}{m_c^4} \psi_{12}\sim\frac{\Lambda_{\rm QCD}^3m_b}{m_c^4}\psi_{12};\nonumber\\
         \\
         \label{x33c}
      x_3^2\Lambda_{\rm QCD}^2\psi_{23}:&&
      \Lambda_{\rm QCD}^2\psi_{23}\frac{8(\kappa_3-q')^2}
         {\left[m_s^2-(\kappa_3-q')^2\right]^2}
\big|_{\kappa_1=\omega_1 p}
\sim \frac{\Lambda_{\rm QCD}^2\omega_3 M_B^2}{(\omega_3 M_B^2)^2}\psi_{23}
\sim\frac{\Lambda_{\rm QCD}}{m_b}\psi_{23}; \nonumber\\
         \\
         \label{x13c}
 x_1 x_3\Lambda_{\rm QCD}^2\psi_{13}:&&
 \frac{2u(1-u)(\kappa_1^\mu -q^\mu \tau)}{m_c^2-u(1-u)[(\kappa_1-q)^2\tau+\kappa_1^2(1-\tau)]}
 \frac{2(\kappa_3-q')^\mu}{m_s^2-(\kappa_3-q')^2}\bigg|_{\kappa_{1,3}=\omega_{1,3}p}
\sim \Lambda_{\rm QCD}\frac{q'q}{m_c^2 m_b}\psi_{13}
   \sim \frac{\Lambda_{\rm QCD} m_b}{m_c^2}\psi_{13}.\nonumber\\
\end{eqnarray}
Taking into account the adopted scaling, $\Lambda\ll m_c\ll m_b$, and $\Lambda m_b/m_c^2\simeq 1$, we see that 
the factors describing the relative contributions of powers of $x_1^2$ and $x_3^2$ are small (i.e.
the dominant contribution comes from the region where $x_1$ and $x_3$ are on the light cone,
$x_1^2=0$ and $x_3^2=0$), whereas, 
as expected from the general arguments, each term of the form $(x_1-x_3)^{2n}$ in $\Psi_p(x_1,x_3)$
leads to the contribution to $A(p,q)$ of the same order as $A_{\psi_0}(p,q)$. 
So, the knowledge of the full functional dependence of $\Psi_p(x_1,x_3)$ on the variable $(x_1-x_3)^2$
is essential for a proper resummation of large $\Lambda_{\rm QCD}m_b/m_c^2$ corrections to the amplitude $A(p,q)$
of FCNC $B$-decay. 

\noindent
$\bullet$ We would like to remark on what happens if one considers the contribution of the light $u$-quark instead of the
$c$-quark in the loop: 
First, in this case one cannot set $q^2=0$ but has to keep $q^2\le -1$ GeV$^2$ to
validate the perturbative calculation of the $u$-quark loop. 
Second, the scaling relations change 
\begin{eqnarray}
  \label{x11u}
x_1^2\Lambda_{\rm QCD}^2\psi_{12}:&&\sim\frac{\Lambda_{\rm QCD}}{m_b}\psi_{12}; \\
\label{x33u}
      x_3^2\Lambda_{\rm QCD}^2\psi_{23}:&&\sim\frac{\Lambda_{\rm QCD}}{m_b}\psi_{23}; \\
\label{x13u}
 x_1 x_3\Lambda_{\rm QCD}^2\psi_{13}:&& \sim \psi_{13}. 
\end{eqnarray}
Obviously, the statement that all $(x_1-x_3)^{2n}$ terms provide the $O(1)$ contributions applies 
to both cases of the $c$ and the $u$-quarks in the triangle diagram.

\noindent$\bullet$
Finally, let us emphasize the following feature of the $B\to j_1 j_2$ amplitude:
If in Eq.~(\ref{x13c}) we keep the leading term only and neglect all corrections  
$O(\Lambda_{\rm QCD}/m_b)$, then the contributions of $\psi_{13}$ simplifies considerably 
and takes the form similar to the contribution of $\psi_0$ (\ref{Aqp}). Moreover,
the contribution of the full 3BS Eq.~(\ref{3da}), including all powers of $(x_1-x_3)^{2n}$
($\psi^{(1)}_{13}\equiv \psi_{13}$),
\begin{eqnarray}
\label{3da1}
\Psi_p(x_1,x_3)=
\int d \omega_1 d\omega_3 e^{-i \omega_1 x_1 p -i \omega_3 x_3 p}
\left[\psi_0(\omega_1,\omega_3)+\sum\limits_{n=1} \psi^{(n)}_{13}(\omega_1,\omega_3)x_{13}^{2n}+O(x_1^2,x_3^2)\right],
\end{eqnarray}
to a FCNC $B\to j_1j_2$ amplitude may be written with $O(\Lambda_{\rm QCD}/m_b)$ accuracy
in a simple form \cite{wang2022private}: 
\begin{eqnarray}
\label{Aqpfull}
A(q,p)=\int\limits_0^1 d\omega_1 \int\limits_0^{1-\omega_1} d\omega_3 \,\psi_{\rm eff}(\omega_1,\omega_3)
\Gamma_{cc}\left(\omega_1 p, q \right)D_s\left((q'-\omega_3 p)^2\right), 
\end{eqnarray}
with
\begin{eqnarray}
\label{psieff}
\psi_{\rm eff}(\omega_1,\omega_3)=\psi_{0}(\omega_1,\omega_3)
+\frac{4}{M_B^2}\frac{\partial^2}{\partial\omega_1\partial\omega_3}\psi_{13}(\omega_1,\omega_3)+\dots. 
\end{eqnarray}
Here the dots stand for the contributions of higher functions $\psi_{13}^{(n)}$, $n\ge 2$.
The expression (\ref{psieff}) as well as the contributions of $\psi_{13}^{(n)}$ 
may be easily obtained making use of the relation \cite{wang2022private}:
\begin{eqnarray}
\label{constraint}
x_{1}x_{3}=\frac{2}{M_B^2}x_1p\,x_3 p. 
\end{eqnarray}
The latter relation is valid to $O(\Lambda_{\rm QCD}/m_b)$ accuracy as soon as $x_1^2$ and $x_3^2$ are near the LC, 
$x_1^2=O(1/\Lambda_{\rm QCD}m_b)$ and $x_3^2=O(1/\Lambda_{\rm QCD}m_b)$.

As said above, in general, the invariant amplitudes appearing in 3BS are functions of 5 independent variables:
$x_1 p$, $x_3 p$, $x_1^2$, $x_3^2$, $(x_1-x_3)^2$. However, if $x_1^2=0$ and $x_3^2=0$, the variable
$(x_1-x_3)^2$ is not an independent variable anymore and is reduced to the combination of the variables $x_1 p$ and $x_3 p$,
Eq.~(\ref{constraint}). 

\section{Conclusions}
We performed a detailed comparison of the contributions coming from
3-particle BS amplitude of the $B$-meson
\begin{eqnarray}
\langle 0|\bar s(x)G(z)b(0)|B(p)\rangle 
\end{eqnarray}
to
(i) $B$-meson weak decay form factor that describes the CC $B$-decay amplitude and to
(ii) non-factorizable part of FCNC $B$-decay amplitude. 
Both amplitudes are related to the same diagram of the form factor topology with, however, a different location
of the heavy-quark field:
In the CC amplitude, the heavy $b$-quark
is located at the end point of the line along which fast light quarks propagate in the diagram,
whereas in the FCNC amplitude the heavy $b$-quark hits the middle of this line. 
As the result, the dominant contributions to these two amplitudes come from different 3BS configurations:
\begin{itemize}
\item
The dominant contribution to a CC amplitude comes from the collinear configuration when both vertices $x$ and $z$
lie along the same light-cone direction $z_\mu=u x_\mu$, $x^2=0$ and $z^2=0$. Therefore, 
a collinear 3BS $\langle 0|\bar s(x)G(u x)b(0)|B(p)\rangle$
is sufficient for the calculation of the 3BS correction to the form factor.
\item
The dominant contribution to a FCNC amplitude comes from a different configuration when both vertices $x$ and $z$
lie on the light cone, but along the different light-cone directions such that $z^2=x^2=0$, but $xz\ne 0$.
Therefore, a non-collinear 3BS of the $B$-meson, $\langle 0|\bar s(x)G(z)b(0)|B(p)\rangle$, with $x^2=z^2=0$, but
$x-z)^2\ne 0$, is necessary for a reliable calculation of the 3BS correction to the FCNC form factor.
\end{itemize}
We point out that a simple physics picture lies beyond these results:
One considers the $B$-meson rest frame; in this rest frame the $b$-quark is almost at rest.

In a CC decay, the $b$ quark decays in a fast lepton pair with momentum $q$ and a fast light quark which 
moves, say, along the $+$ light-cone direction. At the point $z$ it is hit by a soft gluon
and continues to move practically along the same direction before it reaches the point $x$ where it emits
the momentum $q'$. So we come to the well-known result that the 3BS correction to the $B$-decay CC form factor
is dominated by the collinear light-cone configuration $x^2=0$, $z^2=0$, and $z_\mu =u x_\mu$. 

In a FCNC amplitude the situation is different: a resting $b$-quark emits a fast $s$-quark in one space direction
and a fast pair of charmed quarks\footnote{Obviously, the picture does not change if $c$-quark in the triangle
is replaced by $u$-quark.} in an opposite space direction. If we translate this into light-cone directions
and assign the direction of the $s$ quark as $(+)$, then the $c$-quark pair moves along the $(-)$ LC direction.
At point $x$ the $s$-quark emits the momentum $q'$. The point $x$ thus lies on the LC along its $(-)$ direction.
The fast $c$-quark pair is hit by the soft gluon at the point $z$ and continues to move up to the point $z'$
where it emits the momentum $q$. Both $z$ and $z'$ lie along the $(+)$ LC direction.
We see that $x^2=z^2=0$, but, in general, $(x-z)^2\ne 0$.

Notice, that this argument does not say yet that all $O\left((x-z)^{2n}\right)$ terms in 3BS of
the $B$-meson lead to $O(1)$ contributions compared to the contribution of the $\psi_0$ term.

The analysis of Section \ref{sect:fcnc} showed that for the case of the $c$-quark in the loop, the dominant contributions 
to the FCNC amplitude come from the region $x^2=z^2=0$, and $\Lambda_{\rm QCD}^2 (x-z)^2\simeq {\Lambda_{\rm QCD}m_b}/{m_c^2}=O(1)$.
In this case, all $O\left((x-z)^{2n}\right)$ terms in the 3BS of the $B$-meson lead to $O(1)$ contributions
compared to the contribution of the $\psi_0$ term. The same result holds for the light quark in the triangle instead of the
$c$-quark.

In conclusion, let us recall that the idea of going from local OPE for the FCNC amplitude to a non-local OPE
was motivated by the necessity to sum up large $(\Lambda_{\rm QCD}m_b/m_c^2)^n$ corrections to the FCNC amplitude.
However, keeping only the collinear LC part of the 3BS of the $B$-meson and
neglecting all terms of the order $(x-z)^{2n}$ leads to the resummation of a part of these large $(\Lambda_{\rm QCD}m_b/m_c^2)^n$
corrections, whereas another source of the corrections of the same order of magnitude remains unaccounted. 
So, the full dependence of 3BS of the $B$-meson on the variable $(x-z)^2$ is necessary to sum properly the
$(\Lambda_{\rm QCD}m_b/m_c^2)^n$ corrections. 

\vspace{.2cm}
\noindent
{\bf\it Acknowledgments.}
I am grateful to I.~Belov, A.~Berezhnoy, M.~Ferre, E.~Kou, O.~Nachtmann, and H.~Sazdjian for fruitful
discussions, to E.~Thommes and G.~Wolschin for their kind hospitality, and
to the AvH-Stiftung for support. My special thanks are due to Q.~Qin, Y.-L.~Shen, C.~Wang and
Y.-M.~Wang for illuminating comments concerning their paper \cite{wang2022}. 

\end{document}